\documentclass[12pt]{article}
\usepackage[pdftex]{graphicx}
\usepackage{wrapfig}
\usepackage{latexsym}
\usepackage[pdftex]{color}
\usepackage{pifont}
\usepackage{epstopdf}
\usepackage{tikz}
\usepackage{mathtools}
\usetikzlibrary{decorations.pathmorphing,patterns}

\newcommand{\bmp}[2][t]{\begin{minipage}[#1]{#2}}
\newcommand{\emp}{\end{minipage}}

\usepackage{amsmath}
\usepackage{amssymb}
\usepackage{textcomp}
\usepackage{tcolorbox}
\usepackage[title]{appendix}

\usepackage{sidecap,caption}

\usepackage[square,numbers,sort&compress]{natbib}

\newcommand\black{\color{black}}

\newcommand{\bblubox}[1]{\begin{tcolorbox}[colframe=blue!75!white,title=#1]}
\newcommand{\eblubox}{\end{tcolorbox}}

\begin{document}

\title{Kinetic Membrane Model of Outer Hair Cells {I:}\\  Motile Elements with Two Conformational States}

\author{K. H. Iwasa\\
\normalsize{Bldg.\ 35A, Rm 1F121A}\\
\normalsize{NIDCD, National Institutes of Health}\\
\normalsize{Bethesda, Maryland 20892}
}

\date{\small{date: 5 March 2020}}
\maketitle

\noindent running title: \emph{Membrane Model of Outer Hair Cells} \\ 
keywords: \emph{piezoelectricity, SLC26A5, equation of motion, turgor pressure}
\abstract{The effectiveness of outer hair cells (OHCs) in amplifying the motion of the organ of Corti, and thereby contributing to the sensitivity of mammalian hearing, depends on the mechanical power output of these cells. Electromechanical coupling in OHCs, which enables these cells to convert electrical energy into mechanical energy, has been analyzed in detail using isolated cells using primarily static membrane models. In the preceding reports, mechanical output of OHC was evaluated by developing a kinetic theory based on a simplified one-dimensional (1D) model for OHCs. Here such a kinetic description of OHCs is extended by using the membrane model, which has been used for analyzing in vitro experiments. The present theory predicts, for systems without inertial load, that elastic load enhances positive shift of voltage dependence of the membrane capacitance due to turgor pressure. For systems with inertia, mechanical power output also depends on turgor pressure. The maximal power output is, however, similar to the previous prediction  of up to $\sim$10 fW  based on the 1D model.
}

\pagebreak

\section{Introduction}
Outer hair cell (OHC) motility is essential for the sensitivity, frequency selectivity, and the dynamic range of the mammalian ear \cite{lib-zuo2002}. This motility is piezoelectric~\cite{i1993,mh1994,ga1994,doi2002}, based on a membrane protein SLC26A5 (prestin)~\cite{zshlmd2000}. This motility is driven by the receptor potential generated by the sensory hair bundle of the each cell. Even though the biological role of this motility has been confirmed by replacing it with its nonfunctional mutants \cite{Dallos2008}, the mechanism, with which OHC motility plays this role, has not been fully clarified. 

One of the problems is  the  so called ``RC time constant problem'' due to the intrinsic electric circuit of the cell, which is expected to heavily attenuate the receptor potential at the operating frequencies  of hair cells \cite{ha1992}. That is because, for the membrane potential to change, an electric current need to charge up or down the capacitor, consisting of the plasma membrane. To address this puzzle, a possibility has been explored that the cell was primarily driven by the extracellular voltage (cochlear microphonic), the attenuation of which with increasing frequency is less steep \cite{Dallos1995,Mistrik2009}. However, the answer can be sought within the cell itself by examining the resistance (R) and the capacitance (C). Even though the membrane resistance can be overestimated  in  \emph{in vitro} preparations \cite{Johnson2011}, a lower membrane resistance does not lead to larger receptor potential even though it does reduce the RC time constant. Instead, a reduction of the membrane capacitance does increase the receptor potential at frequencies above the cell's RC-corner frequency. While the movement of the motile molecule's charge increases the membrane capacitance of the cell (nonlinear capacitance) under load-free condition, it is sensitive to mechanical load. It was found that elastic load can reduce the membrane capacitance \cite{Iwasa2016}. Moreover, nonlinear capacitance can turn negative and eliminate the membrane capacitance altogether near resonance frequency in the presence of inertial load \cite{Iwasa2017}.

Previous theoretical efforts did address issue of high frequency response \cite{Spector2003,Rabbitt2009} and the effect of elastic load \cite{Rabbitt2009}. However, these efforts did not consider the effect of load on the membrane capacitance, nor inertial load with an exception of a cochlear model \cite{OMaoileidigh2013} until the preceding reports \cite{Iwasa2016,Iwasa2017}, which \black evaluated the effectiveness of OHCs by comparing the power output of OHC with energy dissipation in the subtectorial gap, essential for mechanoreception of the ear, in view of the importance of energy balance involving OHCs \emph{in vivo} \cite{Ramamoorthy2012a,Wang2016}. Power output of OHCs was obtained by constructing the equation of motion of OHC with mechanical load.  Nonetheless, the puzzles have not been fully clarified because these theoretical treatments include various approximations, which could affect the outcome, in addition to the paucity of experimental confirmations.   

Another issue that is important for physiological role of OHCs is the intrinsic transition rates of prestin. Nonlinear capacitance measured using the on-cell mode of patch clamp, which allows applying voltage waveforms without a low pass filter, rolled-off at $\sim$15 kHz \cite{ga1997}  at room temperature or at much lower frequencies \cite{Santos-Sacchi2018}. Cell displacements elicited by voltage waveforms applied through a suction pipette showed  frequency responses ranging from 6 to 80 kHz \cite{fhg1999}  or  lower at 8.8 kHz \cite{Santos-Sacchi2018}. 
Current noise spectrum from sealed patch formed on OHC showed relaxation time constant of 35 kHz, different from the value $\sim$15 kHz for nonlinear capacitance, presumably reflecting a different mode of motion \cite{Dong2000}. These observations indicate that those time constants reflect the mechanical relaxation time of each system rather than the intrinsic time constant of the motile molecules.

Those earlier data are contradicted by the most recent reports from Santos-Sacchi's group on the frequency dependence of nonlinear capacitance and length changes in the whole-cell mode, by performing capacitance compensation after recording \cite{SantosSacchi2019}. The intrinsic gating frequency that they inferred is as low as 3 kHz for guinea pigs, deduced only after capacitance compensation. These authors also measured the frequency dependence of nonlinear capacitance of sealed membrane patches \cite{SantosSacchi2019}. However, charge movements in a sealed membrane patch could be affected by the viscoelastic frequency as earlier experiments have shown \cite{Dong2000}.

An examination of power production  based on a 1D model of OHC shows, however, that prestin with slow intrinsic gating rate are incapable of overcoming the shear drag between the tectorial membrane and the reticular lamina beyond the gating frequency \cite{Iwasa2020p}.  Since this shear drag is  indispensable for stimulating hair cells \cite{hcmm2000}, such slow gating of prestin contradicts the physiological role of OHCs as well as energy conservation without assuming a significant lateral energy flow, which is unlikely \cite{Wang2016a}. 

The present paper assume that the intrinsic rate is faster than the auditory range, to follow up the preceding reports based on a one-dimensional model \cite{Iwasa2016,Iwasa2017}. It has two specific goals. One is to formulate a kinetic theory by extending the membrane model, which is more physical and has been used for analyzing quasi-static \emph{in vitro} experiments  yet simpler than shell model with more parameters \cite{sbp1999}.   This enables us to examine the limitations and accuracies of the 1D model \cite{Iwasa2016,Iwasa2017}.  It also allows to describe the effect of turgor pressure. The other goal is to prepare for extension to a more complex theory, in which the motile element has multiple states so that the effect of anion binding can be described \cite{oliv-fakl2001,tga1995,Homma2011,Santos-Sacchi2014}. Such an extension will be presented in the next report. In addition, amplifier gain of OHCs is discussed.

\section{The system}

\begin{SCfigure}
\includegraphics[width=0.45\linewidth]{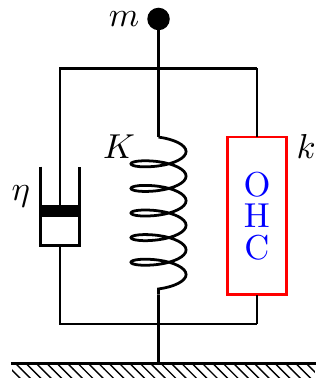} 
\captionsetup{width=1.1\linewidth}
\caption{\small{An OHC  (red rectangle) is subjected to mechanical load, consisting of viscous $\eta$ (left), elastic $K$ (middle), and inertial  $m$ (top) components. The stiffness of the cell is $k$. In its lateral membrane the cell has a membrane protein, which undergoes conformational transitions between two states. The density of the proten is $n$ and the conformational transitions involve transfer of electric displacement $q$ across the membrane and mechanical displacements of $a_z$ and $a_c$ respectively in the axial and circumferential directions (Fig.\ \ref{fig:cell}).}}
\label{fig:load}
\end{SCfigure}

Here we consider a system, in which an OHC has mechanical load, consisting of viscous, elastic, and inertial components (Fig.\ \ref{fig:load}). It is not intended to simulate the organ of Corti, where OHCs are localized. Instead, this model system is intended to provide an OHC with a simplest possible environment so that its performance can be examined. The biological role of this cell could be inferred from this examination.

Let us assume that the cell maintains its characteristic cylindrical shape. Such a simplifying assumption could be justified for low frequencies, where inertial force is not significant compared with elastic force within the cell. The limit of the validity is examined later in Discussion, after experimental values for material properties are provided.

In addition, we consider movement of this cell in response to small fast changes in the membrane potential, from an equilibrium condition, ignoring metabolic processes that maintained the physiological condition. Changes in turgor pressure, if present, is assumed gradual and therefore has only a modulatory role. For this reason, the volume of the cell is assumed constant during voltage changes.

\subsection{Electrical connectivity} 
Here we initially assume that the membrane potential of the cell is controlled by an extracellular electrode and an intracellular electrode with low impedance to facilitate the evaluation of the membrane capacitance. Later on, for evaluating power output of an OHC, the mechanotransducers at the hair bundle will be introduced because the receptor potential that drive the motile element in the OHCs depends on the membrane capacitance as well as the current source.

\subsection{Static properties of OHC}
First, consider an OHC in equilibrium. The shape of an OHCs is approximated by an elastic cylinder of radius $r$ and length $L$.
\begin{SCfigure}
\centering
\includegraphics[width=0.5\linewidth]{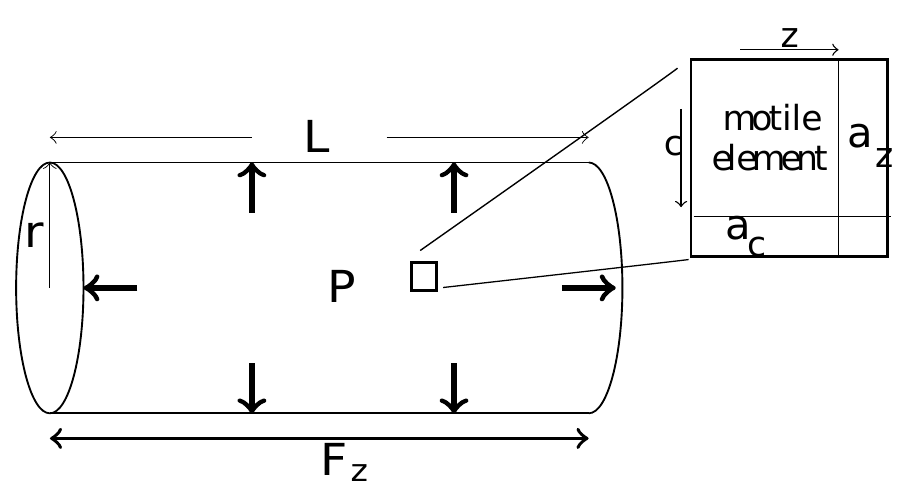} 
\captionsetup{width=1.1\linewidth}
\caption{\small{The membrane model of OHC. The cell body of an OHC is approximated by a cylinder of radius $r$ and length $L$. Motile elements (upper right), which undergo area changes, $a_z$ in the axial direction and $a_c$ in the circumferential direction, are uniformly embedded in the cylindrical plasma membrane at density $n$.
}}
\label{fig:cell}
\end{SCfigure}

Small displacements of the cell can be described by constitutive equations \cite{i2001},
\begin{subequations}
\label{eq:constitutive}
\begin{align}
 d_1\epsilon_z'+g\epsilon_c'&=f_z+\frac 1 2 rP,\\
g\epsilon_z'+d_2\epsilon_c'&= rP,
\end{align}
\end{subequations}
where $\epsilon_z'$ and  $\epsilon_c'$ are respectively small elastic strains of the membrane in the axial direction and the circumferential direction. The quantities $d_1$, $d_2$, and $g$ are elastic moduli of the membrane; $P$ the internal pressure and $f_z$ the axial tension due to an external force $F_z$, which can be expressed as $2\pi r f_z$. These elastic moduli assume orthotropy \cite{ts1995}. 
 
\subsection{Axial stiffness}
The material stiffness $k$ in the axial direction and the axial elastic modulus $\kappa$ of the cell are expressed as
\begin{align}\label{eq:k_kappa}
 k&=\frac{2\pi r}{L_0}\kappa, 
\end{align}
where $L_0$ is the resting length of the cell. The axial modulus $\kappa$ is obtained as $\kappa=f_z/\epsilon_z'$ under constant volume condition $\epsilon_v'=\epsilon_z'+2\epsilon_c'=\mathrm{const.}$, eliminating $\epsilon_c'$ and $rP$ from the constitutive equations (Eq.\ \ref{eq:constitutive}). It can be expressed as (See Appendix \ref{apx:eqs}).
\begin{align}  \label{eq:kappa}
 \kappa&=d_1-g+d_2/4.
\end{align}
This axial stiffness of the cell is material stiffness without motile element, which would reduce the stiffness in a manner similar to ``gating compliance.'' \cite{hh1988,i2000}

\section{Motile element} 
Here we assume that each motile element undergoes transition between two states, compact (C) and extended (E). In the following, the fraction of the state is expressed in italic. \footnote{The fraction of state C is used as the variable to describe the state of the motile element in this treatment so that this treatment can be extended to a system with motile elements, which undergo transitions between multiple states.}

\begin{align*} \nonumber
\mathbf{E}(\mathrm{extended})&\qquad \rightleftharpoons \qquad\mathbf{C}(\mathrm{compact}). 
\end{align*}
Unlike conventional description of molecular transitions, we do not assume that transitions between these states take place in accordance with ``intrinsic'' transition rates. Instead, we assume that transitions between these states are determined by the mechanical constraints in a manner similar to piezoelectricity.

To incorporate the motile element,  let us assume that the total strain in the axial direction $\epsilon_z$ and the one in the circumferential direction $\epsilon_c$ consist of elastic components ($\epsilon_z'$ and $\epsilon_c'$) and the contribution of motile elements. Each motile element undergoes electrical displacement $q$ and mechanical displacements $a_z$ and $a_c$ (see Fig.\ \ref{fig:cell}) during the transition from state E to state C. If the motile elements are uniformly distributed in the lateral membrane at density $n$, the total strains in the two directions are then respectively expressed by \cite{i2001}
\begin{subequations}
\label{eq:tot_strains}
\begin{align} 
 \epsilon_z&=\epsilon_z'+a_znC,\\ 
 \epsilon_c&=\epsilon_c'+a_cnC.
 \end{align}
 \end{subequations}

The constitutive equations are then re-written by
\begin{subequations}
\label{eq:total_strains}
\begin{align}
d_1\epsilon_z+g\epsilon_c-(a_zd_1+a_cg)nC&=f_z+\frac 1 2 rP,\\
g\epsilon_z+d_2\epsilon_c-(a_zg+a_cd_2)nC&= rP.
\end{align}
\end{subequations}
Notice that the internal pressure $P$ now consists of two components, dependent on the activity of the motile element, and independent of it.

In the presence of external elastic load $K$, axial  tension  $f_z$ can be expressed by 
\begin{align}\label{eq:zfz1}
 f_z=-K_e\epsilon_z,
\end{align}
where $K_e=KL_0/(2\pi r)$. Eqs.\ \ref{eq:total_strains} and \ref{eq:zfz1} lead to
\begin{align}\label{eq:eps_z1}
\epsilon_z&=-\frac{AnC-\mu \epsilon_v}{2(\kappa+K_e)},
\end{align}
where $A=-(2d_1-g)a_z+2\mu a_c$ and $\mu=d_2/2-g$ (See Appendix \ref{apx:eqs}). The quantity $\epsilon_v$ is the volume strain, which can be expressed as $\epsilon_z+2\epsilon_c$ for the cylindrical cell for small strains. 

Since the displacement of the cell that we are interested in is in the auditory frequency range, it is reasonable to assume the cell volume is constant. For this reason, we can regard $\epsilon_v$ as a variable representative of turgor pressure, which does not depend on the activities of the motile elements and can change only slowly responding to metabolic activity or osmotic pressure. 

The axial displacement $z(=\!L_0\epsilon_z)$ of the cell and the total charge  of the motile elements  $Q$ can be expressed respectively using $C$, the fraction of state C by \footnote{Since the fraction of state E was used in the previous reports, the resulting equations have the negative sign on $A$ as well as on $q$. This alteration of the notations facilitates an extension of the present treatment into a system, where the motile elements undergo multiple state transitions.}
\begin{subequations}
\begin{align}\label{eq:z2} 
 z&=-\frac{\pi r}{k+K}\cdot(AnC-\mu \epsilon_v),\\ \label{eq:Q}
 Q&=-qNC,
\end{align}
\end{subequations}
where $N$ the total number of motile elements, i.e. $N=2\pi rL_0 n$.

These equations are similar to those for the one-dimensional model described earlier. 
The difference is that the previous equations did not have turgor pressure dependence, 
which is expressed by the second term on the RHS of Eq.\ \ref{eq:z2}. 

\subsection{Boltzmann distribution}
The fraction of state C in equilibrium should be given by a Boltzmann function
\begin{align}
 C_\infty&=\exp[-\beta \Delta G_m]/(1+\exp[-\beta\Delta G_m]),\\ 
\mathrm{with}  \quad
\Delta G_m&=-q(V-V_{0})-a_zf_z-(a_z/2+a_c)rP
\label{eq:Gm}
\end{align}
where  $\Delta G_m$ is the energy difference (of state C from state E),   $V$  the membrane potential, $\beta=1/k_BT$ with Boltzmann's constant $k_B$, and the temperature $T$.  The voltage $V_0$ determines the operating point.  By substituting $f_z$ and $rP$, we obtain (See Appendix \ref{apx:eqs})
\begin{align}\label{eq:DGm} 
 \Delta G_m=-q(V-V_0)+\frac{1}{4\kappa}[(A^2\hat{K}+\varphi a^2)nC-(\mu A\hat{K}+\varphi a)\epsilon_v],
\end{align}
where shorthand notations are
\begin{subequations}
\begin{align}
a&=-(a_z+2a_c),\\
\hat K&=K_e/(\kappa+K_e)=K/(k+K),\\
\varphi&=d_1d_2-g^2.
\end{align}
\end{subequations}
The volume $\epsilon_v$ is due to turgor pressure $P_t$. Under the condition that the motile element is not active, $P_t=\epsilon_v\varphi/(2\kappa r)$.

\section{Equation of motion}
If $C=C_\infty$, i.e. the distribution of motor states are in equilibrium, the cell does not move.  Suppose the membrane potential $V$ changes abruptly, the cell exerts force $k\cdot \pi r An(C_\infty-C)$ because a change $\Delta C$ in $C$ gives rise to length change $\pi r An\Delta C$ of the cell due to the cylindrical geometry if $\epsilon_c$ is kept constant. Here $k$ is the material stiffness of the cell defined by Eq.\ \ref{eq:kappa}). The equation of motion can be expressed as
\begin{align}\label{eq:eomz}
m\frac{d^2z}{dt^2}+\eta\frac{dz}{dt}=\pi r kAn(C_\infty-C),
\end{align}
where $C_\infty$ is the value of $C$,  which satisfies Boltzmann distribution,  for the given condition at time $t$.  With the aid of Eq.\ \ref{eq:z2}, this equation can be re-written in the form 
\begin{align}
 m\frac{d^2C}{dt^2}+\eta\frac{dC}{dt}=(k+K)(C_\infty-C),
 \label{eq:eom2}
\end{align}
which is intuitive, considering that $z$ is proportional to $C$.
This equation determines the rates of transitions between the two states because $C$ is the only variable, given $C+E=1$.  

This is the same equation derived earlier based on the one-dimensional model \cite{Iwasa2016} 
except that the variable $C$ in this membrane model depends on a larger number of factors, 
including turgor pressure. The property of the OHC as expressed by the equation cannot 
be rendered as a simple combination, either series or parallel, of an elastic element and a 
displacement element. 

Notice that,  the presence is the first term in Eq.\ \ref{eq:eomz} (and also in Eq.\ \ref{eq:eom2}) is an assumption.  In the absence of the inertial term, Eq.\ \ref{eq:eomz} has the typical form of a relaxation equation.  Even though this equation appears reasonable at the moment $dz/dt=0$,  the presence of the inertial term may not be solidly justified. However,  it will be shown later that this equation with the inertial term can be justified, being  equivalent to the the equation that describes piezoelectric resonance if the deviation from equilibrium is small  (See Discussion \ref{subsec:piezo}).

\subsection{Small harmonic perturbation}
Since one of the main functions of OHCs is to amplify small signal, the response of OHCs to small harmonic stimulation is of special interest. Assume that the voltage consists of two parts, a constant term $\langle V\rangle$ and small sinusoidal component  with angular frequency $\omega$ and amplitude $v$:
\begin{align*}
 V(t) &=\langle V\rangle+ v \exp[i\omega t].
\end{align*}
Then $C$ and $C_\infty$ should also have two corresponding components
\begin{align}
 C(t)&=\langle C\rangle+ c \exp[i\omega t],\\
 C_\infty(t)&=\langle C_\infty\rangle+ c_\infty \exp[i\omega t],
\end{align}
and the fist-order terms of the equation of motion turns into
\begin{align}
\label{eq:eom}
 [-(\omega/\omega_r)^2+i\omega/\omega_\eta+1]c=c_\infty,
\end{align}
with $\omega_r^2=(k+K)/m$, $\omega_\eta=(k+K)/\eta$, and
\begin{align}
 c_\infty=\gamma[qv+\frac{nc}{2\kappa}(A^2\hat{K}+\varphi a^2)],
 \label{eq:c_inf}
\end{align}
with $\gamma=\beta \langle C\rangle(1-\langle C\rangle)$. Thus quantity $c$ obeys the equation
\begin{align}
 [-(\omega/\omega_r)^2+i\omega/\omega_\eta+\alpha^2]c=\gamma qv,
 \label{eq:cmemb_eq}
\end{align}
with $\alpha^2=1+\gamma n(A^2\hat{K}+\varphi a^2)/(2\kappa)$.

\subsubsection{Nonlinear capacitance}
If we express corresponding changes in a similar manner, the charge variable can be expressed as $Q=nq\langle C\rangle+ Nqc\exp[i\omega t]$, and nonlinear capacitance $C_{nl}$ is given by
\begin{align} \nonumber
 C_{nl}&= Re[Nqc/ v],\\
 &=Re\left[\frac{\gamma Nq^2}{\alpha^2-\overline\omega^2+i\overline\omega/\overline\omega_\eta}\right]
\end{align}
where $Re[...]$ represents the real part  because capacitance is charge transfer synchronous to voltage changes \cite{Iwasa2016}. Here shorthand notations are introduced:  $\overline\omega=\omega/\omega_r, \;\overline\omega_\eta=\omega_\eta/\omega_r$. The axial displacement $z$ can be evaluated using Eqs.\ (\ref{eq:z2}) and (\ref{eq:eom}).

\section{Comparison with one-dimensional model} 
How the predictions of the membrane model differs from 1D models? In the following, the 1D model that was previously introduced \cite{Iwasa2016,Iwasa2017} is briefly restated to facilitate the comparison.

\subsection{One-dimensional model}
A 1D model has a single parameter $k$ for the cell's elasticity and a single parameter $a_1$ for mechanical changes of the motile elements. For length changes $z$ and charge transfer $Q$, the equation that correspond to Eqs.\ \ref{eq:z2} can be written down~\cite{Iwasa2016}, \footnote{The previous treatments used a variable $P_\ell$, which corresponds to $P_\ell=1-C$. Thus the signs of $a_1$ and $q$ are reversed in Eqs.\  \ref{eq:zQ1} to keep the the signs of $q$ and $a$ the same as the previous treatments. In addition, $\tilde K=k \hat K$.}
\begin{subequations}
\label{eq:zQ1}
\begin{align}\label{eq:z1}
z&=\frac{-a_1kNC}{k+K}\\
Q&=-qNC,
\end{align}
\end{subequations}
where $C$ represents the fraction of the compact state, as in the membrane model. The comparison of Eqs.\ \ref{eq:z2} and \ref{eq:z1} suggest $a_1$ corresponds to $A$. The free energy difference $\Delta G_1$ for the 1D model can be expressed by~\cite{Iwasa2016,Iwasa2017}
\begin{align}
 \Delta G_1=q(V-V_{1})+ a_1^2 N k\hat K C.
 \label{eq:G1}
\end{align}

This energy difference $\Delta G_1$ determines $C_\infty=\exp[-\beta\Delta G_1]/(1+\exp[-\beta\Delta G_1])$ and contributes to the factor $\gamma_1=\beta \langle C\rangle(1-\langle C\rangle)$ in the equation of motion. The equation for $c$ for the 1D model is
\begin{align}
  [-\overline\omega^2+i\overline\omega/\overline\omega_\eta+\alpha_1^2]c=\gamma_1 qv,
 \label{eq:c1d_eq}
\end{align}
with $\alpha_1^2=1+\gamma_1 Na_1^2k\hat K$.

\subsubsection{Nonlinear capacitance}

Since charge transfer is $Nqc$, nonlinear capacitance $C_{1nl}$ is expressed by
\begin{align}
  C_{1nl}
 =Re\left[\frac{\gamma_1 Nq^2}{\alpha_1^2-\overline\omega^2+i\overline\omega/\overline\omega_\eta}\right]
\end{align}

\subsection{Correspondence between the two models}
The charge transfer $q$ is identical in the two models. The density $n$ of the membrane model is related to N by $n=N/(2\pi r L_0)$. The relationship between the mechanical factors can be obtained by comparing Eq.\ \ref{eq:z1} with Eq.\ \ref{eq:z2}.  These two equations together with Eq.\ \ref{eq:k_kappa} lead to the expression of unitary length change $a_1$
\begin{align}\label{eq:a1}
 a_1=\frac{A}{4\pi r\kappa}.
\end{align}

\begin{table}[h!]
\caption{\small{Correspondence of parameters in the membrane model and in the 1D model. }}
\begin{center}
\begin{tabular}{c|c}
\hline\hline
1D model  & membrane model  \\
\hline
$a_1$ & $A/(4\pi r\kappa)$ \\
$k$ &$2\pi r\kappa/L_0$   \\
$N$ & $2\pi r L_0 n$ \\
$\Delta G_1$ & $\Delta G_m$ \\
$\alpha_1^2=1+\gamma_1 Na_1^2k\hat K$ & $\alpha^2=1+\gamma n(A^2\hat{K}+\varphi a^2)/(2\kappa)$ \\
\hline
\end{tabular}
\end{center}
\label{tab:correspondence}
\end{table}%

\subsubsection{Dependence on elastic load and static turgor pressure}

It might appear possible to formally replace $C$ with $C-C_0$ in the 1D model to introduce the effect of turgor pressure. If we sort the resulting equations for the dependence on $\hat K$, we may find a correspondence
\begin{subequations}
\begin{align} \label{eq:v1}
qV_1&\Longleftrightarrow qV_{0}-\frac{\varphi a}{4\kappa}\epsilon_v,\\ \label{eq:C0}
a_1^2Nk(C-C_0)\hat K&\Longleftrightarrow \frac{A^2n}{4\kappa}(C+\frac{\mu}{An} \epsilon_v)\hat K,
\end{align}
\end{subequations}
leaving out a term $\varphi a^2nC/(4\kappa)$ in Eq.\ \ref{eq:cmemb_eq}. 

Since $a_1^2Nk=A^2n/(4\kappa)$, Eq.\ \ref{eq:C0} might suggest correspondence $C_0=-\mu \epsilon_v/An$, as if turgor pressure $\epsilon_v$ could be partially introduced through $C_0$. 
Such a modification of Eq.\ \ref{eq:c1d_eq}, however, cannot be justified partially because there is no justification for including a term $a_1^2NkC_0\hat K$ in the free energy difference in the 1D model. In addition, $\varphi a^2nC/(4\kappa)$, the term that was left out in the comparison, may not be small. If $\varphi>0$, this term contributes the transition less sharp because it provides an additional negative feedback to the transition. Thus, such a modification predicts unreasonably sharper transitions.

\section{Power output and amplifier gain} 
The above analysis shows that membrane model leads to the equation for $c$, which is similar to the one derived for the 1D model. For example, the relation between $z$ and $c$ is the same for both cases because $a_1kN/(k+K)=An/(\kappa+K_e)$. The difference of the two models originates only from $\gamma$ and $\alpha^2$, which respectively differ from their counterpart $\gamma_1$ and $\alpha_1^2$ (Table \ref{tab:correspondence}). For this reason, the expression for the 1D model will be used in the following with separate definitions for $\gamma$ and $\alpha^2$ for the membrane model and their counterparts $\gamma_1$ and $\alpha^2_1$ for the 1D model.

\subsection{Power output} 
Under physiological conditions, energy output from an OHC depends on the receptor potential $v$, which is generated by a relative change $\hat r$ in the hair bundle resistance. This potential depends on the intrinsic circuit property of the cell as well as charge movement due to changes in $c$, which can be expressed by \cite{Iwasa2016}
\begin{align}
 v=\frac{-i_0\hat r+i\omega Nqc}{\sigma+i\omega C_0},
 \label{eq:recpot}
\end{align}
where $i_0$ is the steady state current,  $\sigma$ the conductance of the basolateral membrane,  and $C_0$ the structural membrane capacitance of the hair cell.

The combination of Eqs.\ \ref{eq:cmemb_eq} and \ref{eq:recpot} can be written down in the form
\begin{align}
 \left[-\left(\frac{\omega}{\omega_r}\right)^2+i\omega\left(\frac{1}{\omega_\eta}+\frac{\gamma Nq^2}{\sigma+i\omega C_0}\right)+\alpha^2\right]c=\frac{\gamma i_0 q\hat r}{\sigma+i\omega C_0}.
 \label{eq:c-r_eq}
\end{align}

For relatively high frequency, where ionic currents is overwhelmed by displacement current $\omega C_0$, we obtain
\begin{align}
 c=\frac{\gamma i_0 q\hat r}{i\omega C_0}\cdot\frac{1}{-\overline\omega^2+i\overline\omega/\overline\omega_\eta+\alpha^2+\zeta},
\end{align}
with $\zeta=\gamma Nq^2/C_0$.

Energy output from an OHC has two components. One is elastic energy, $(1/2)kz^2$ per half cycle, which is recovered at the end of a cycle. The other is dissipative energy,  $(1/2)\eta\omega |z|^2$ per half cycle, which results in power output $W$, which is given by
\begin{align}\nonumber
 W(\omega)&=\frac{\eta\omega^2}{2\pi}|z|^2\\ \label{eq:W}
 &=\frac{\eta}{2\pi}\left(\frac{k}{k+K}\right)^2\cdot\frac{1}{C_0^2}\cdot\left|\frac{\gamma a_1Nqi_0\hat r}{-\overline\omega^2+i\overline\omega/\overline\omega_\eta+\alpha^2+\zeta}\right|^2.
\end{align}

Power generation in the membrane model is closely related to the 1D model \cite{Iwasa2016, Iwasa2017}. The only difference is from the expressions of free energy in the Boltzmann function, resulting in choosing the set $\gamma$ and  $\alpha^2$ for the membrane model or $\gamma_1$  and $\alpha_1^2$ for the 1D model. 

\subsection{Amplifier gain} 
Power gain $G$ of an amplifier is the ratio of output power $W(\omega)$ against input power $W_\mathrm{in}(\omega)$, where $W_\mathrm{out}$ is already expressed by Eq.\ \ref{eq:W} for a given value of $\hat r$, the relative change in the apical membrane resistance due to hair bundle stimulation. 

A relative change $\hat r$ in resistance can be associated with a displacement $x$ of hair bundle tip with
$ \hat r=gx$, where $g$ is the sensitivity of the hair bundle. Since the tips of hair bundles are embedded in the tectorial membrane, the shear between the reticular lamina and the tectorial membrane can be represented by $x$ also. If we can assume that the shear of the subtectorial space $x$ is proportional to the displacement $z$ of the displacement of OHC cell body, we can put $x=\lambda z$, where $\lambda$ is a constant. These relationships lead to $x=\hat r \lambda g$, and
\begin{align}
 W_\mathrm{in}(\omega)=\frac{\eta\omega^2}{2\pi}\left|\frac{\hat r}{ \lambda g}\right|^2.
\end{align}
The power gain $G(\omega)$ can then be expressed by 
\begin{subequations}
\begin{align}\label{eq:Gdef}
 G(\omega)&=\frac{2\pi (\lambda g)^2}{\eta\omega^2}\cdot\frac{W(\omega)}{\hat r^2},\\ \label{eq:Gxpli}
 &=\frac{1}{\omega_r^2}\left(\frac{k}{k+K}\right)^2\cdot\left(\frac{\lambda g}{\overline\omega C_0}\right)^2\cdot\left|\frac{\gamma a_1Nqi_0}{-\overline\omega^2+i\overline\omega/\overline\omega_\eta+\alpha^2+\zeta}\right|^2.
\end{align}
\end{subequations}
Eq.\ \ref{eq:Gdef} shows that power gain is a product of two factors. One part depends on the reduced frequency $\overline\omega(=\omega/\omega_r$. The other depends on the mechanical resonance frequency $\omega_r$. For a given value of $\omega_r$, the condition $G(\omega)=1$ may not be satisfied if the mechanical resonance frequency $\omega_r$ is too high. Let $\omega_{r,\max}$ the maximum value of $\omega_r$, which is compatible with  $G(\omega)=1$.

The limiting frequency $\omega_\ell$, which satisfies $G(\omega)=1$, was evaluated in a previous treatment \cite{Iwasa2017}. It has been shown that the frequency $\omega_\ell$ is somewhat higher than $\omega_{r,\max}$ \cite{Iwasa2017}.


Amplifier gain at a given location can be related to the limiting frequency $\omega_\ell$. Let the mechanical resonance frequency of the location of interest be $\omega_{r2}$. Then the amplifier gain $G_2(\omega)$ at that location can be expressed in a manner similar to Eq.\ \ref{eq:Gxpli}, with $\omega_r$ replaced with $\omega_{r2}$ and re-defining $\overline\omega$ as $\omega/\omega_{r2}$ instead of $\omega/\omega_r$.

This comparison suggests $G_2(\omega_2)=(\omega_r/\omega_{r2})^2$ if $\eta_2/\omega_{r2}=\eta/\omega_r$, where $\omega_2$ is the best frequency and $\eta_2$ the drag coefficient of the location. The value of drag coefficient is $\eta$ at the location of the limiting frequency. For a more apical location, where the condition $\omega_{r2}<\omega_r$ holds, the drag coefficient should satisfy $\eta_2\leq\eta$ because the subtectorial gap, which makes a significant contribution to the drag coefficient, is less narrower for a more apical location. This condition makes the resonance peak sharper. For this reason, the amplifier gain at best frequency $\omega_2$ satisfies $G_2(\omega_2)>(\omega_r/\omega_{r2})^2$. 



\subsection{Inertia-free condition} 
In the absence of the inertia term, the power output turns into
\begin{align}
\label{eq:wnores}
 W(\omega)= \frac{\eta k^2}{2\pi(k+K)^2C_0^2}\cdot\frac{(\gamma a_1Nqi_0\hat{r})^2}{(\omega/\omega_\eta)^2+(\alpha^2+\zeta)^2},
\end{align}
which is a monotonically decreasing function of the frequency $\omega$. As Eq.\ \ref{eq:Gdef} indicates, the power gain $G(\omega)$ is steeply decreasing function of the frequency.


\subsection{Near resonance}
For the system with inertia, the power output has a peak
\begin{align}
\label{eq:wmax}
 W^{(max)}\approx \frac{\gamma\zeta( a_1 Ni_0\hat{r})^2\overline\omega_\eta^4}{\alpha^2+\zeta}\cdot\frac{\eta k^2}{2\pi (k+K)^2C_0},
\end{align}
at $\overline\omega^2=\alpha^2+\zeta-1/(2\overline\omega_\eta^2)$. Since $\alpha^2>1$, such a peak exist if $\overline\omega_\eta<1$. However, power production $W$ is a decreasing function of the frequency $\omega$ for overdamped systems, where $\overline\omega_\eta$ is large.

These equations for power production are essentially the same as those for the 1D model, which has been studied previously \cite{Iwasa2016,Iwasa2017}. The difference is in the definition of $\alpha^2$ and $\zeta$ even though these factors are similar.


\begin{table}[h!]
\caption{\small{Parameter values of membrane model. $e$ is the electronic charge. The parameter values of the motile element reflect that the extended state E is taken as the reference. ($\ast$): The experimentally obtained regular capacitance of an OHC is close to estimates based on the geometrical surface area with the standard value $\sim 1\mu$F/cm$^2$ for specific capacitance.}}
\begin{center}
\begin{tabular}{c|c|r|r}
\hline\hline
notation  & definition & value used & refs. \\
\hline
$d_1$ & axial modulus & 0.046 N/m &  \cite{ia1997} \\
$d_2$ & circumferential modulus & 0.068 N/m &  \cite{ia1997} \\
$g$ & cross modulus &  0.046 N/m & \cite{ia1997} \\
$a_z$ & axial area change & $-4.5$ nm$^2$    & \cite{i2001} \\
$a_c$ & circumferential area change & $0.75$ nm$^2$ & \cite{i2001} \\
$q$ & mobile charge & -0.8 $e$ &   \\
$n$ & density of motile element & $9\times 10^{15}$/m$^2$ & \cite{i2001} \\
\hline
$r$ & radius & 5 $\mu$m & \cite{ia1997} \\
$L_0$ & cell length & 25 $\mu$m & \\
$C_0$ & regular capacitance & 10 pF & ($\ast$) \\
\hline
\end{tabular}
\end{center}
\label{tab:param_list}
\end{table}%

\section{Numerical examination}
The membrane model and the 1D model lead to parallel expressions for mechanical and electrical displacements, which in turn lead to nonlinear capacitance and power output. The difference in the two stems from the difference in $\Delta G_m$ and $\Delta G_1$. Since the results of the 1D model have been previously elaborated, our focus is whether or not the membrane model leads to different results, using a set of parameter values that have been experimentally determined. 

\subsection{Nonlinear capacitance and factor $\gamma$ }
\begin{figure}[h!]
\begin{center}
\includegraphics[width=14cm]{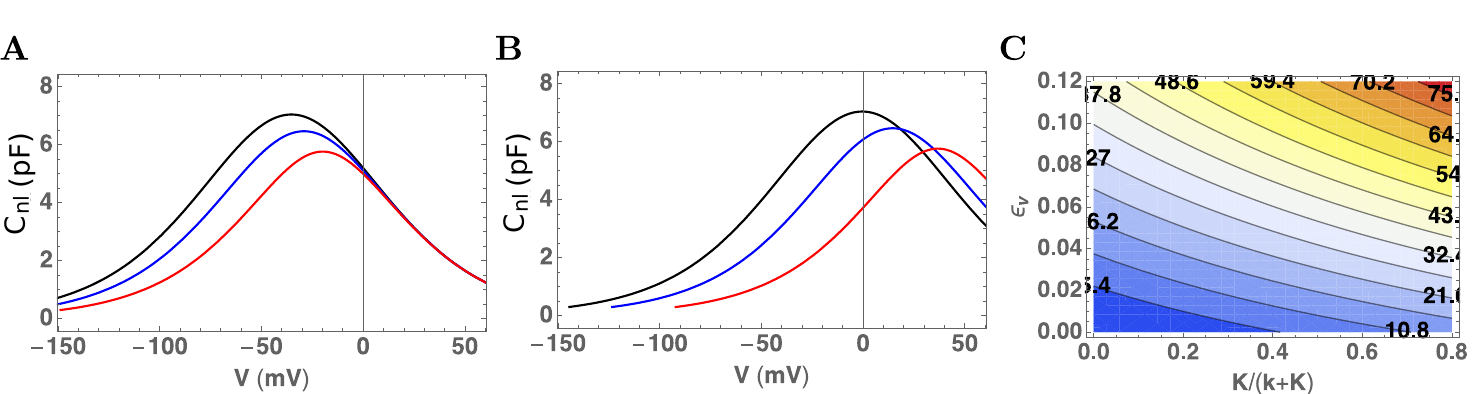} 
\caption{\small{Nonlinear capacitance at low frequency. The volume strain $\epsilon_v$ are 0 for \textbf{A} and 0.10 for \textbf{B}. Traces represent elastic load $K$: 0 (black), 0.01 (blue), and 0.1 (red). C: Contour plot of peak voltage shift. The abscissa: $\hat K(=\!K/(k+K))$. The ordinate axis is volume strain $\epsilon_v$, which represents static turgor pressure.  Voltage shifts are color coded (blue: negative, red: positive) and the values (in mV) are shown in boldface letters in the plot. }}
\label{fig:cap}
\end{center}
\end{figure}

The factor $\gamma$, which contributes to $\alpha^2$ and $\zeta$, is affected by both turgor pressure and external elastic load through $\Delta G_m$ (See Eq.\ \ref{eq:Gm}). This sensitivity is reflected in nonlinear capacitance in the low frequency limit (Fig.\ \ref{fig:cap}). 

Increasing external elastic load broadens the voltage dependence as well as shifts the peak in the positive direction ( Fig.\ \ref{fig:cap}A and B).  An increase in  turgor pressure, represented by $\epsilon_v$, shifts the peak voltage of nonlinear capacitance in the positive direction. This effect of turgor pressure is consistent with earlier studies, both theoretical \cite{i1993,i2001} and experimental \cite{i1993,ks1995,asi2000}. In addition, turgor pressure increases the sensitivity of nonlinear capacitance on the elastic load in both shifting the peak as well as broadening of the dependence (Fig.\ \ref{fig:cap}B). A contour plot of peak voltage shift summarizes the dependence on both turgor pressure and elastic load.

\subsection{Power output}
Power output of an OHC has been described using the 1D model \cite{Iwasa2016,Iwasa2017}.  Here we focus on the issue as to how the predictions of the membrane model compare with those of the 1D model for the given set of the parameters.

\subsubsection{Inertia-free condition}
\begin{figure}[h!]
\begin{center}
 \includegraphics[width=10cm]{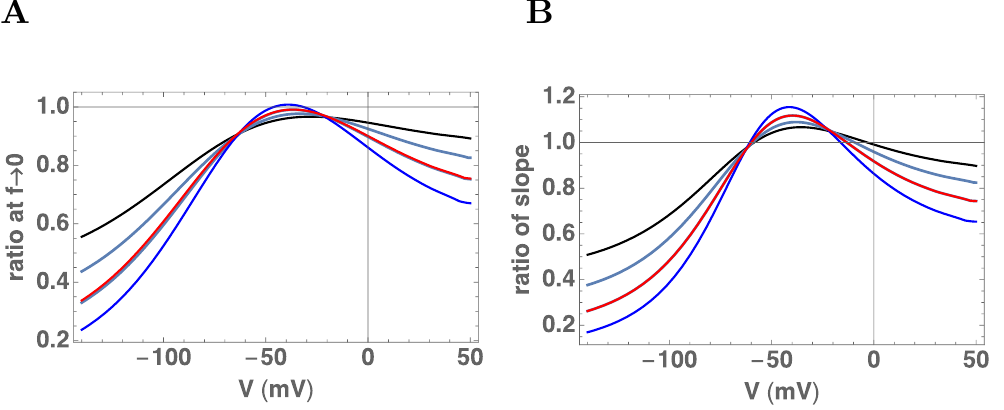} 
\caption{\small{
Ratio of coefficients for power output of the present membrane model to those of the 1D model. A: The ratio of power output of the membrane model to that of 1D model at zero frequency asymptote. B: The ratio at high frequencies. Traces correspond to, $K=0.1 k$ (blue), $0.5 k$ (azure), $k$ (blue), and $2 k$ (red).
}}
\label{fig:noinertia}
\end{center}
\end{figure}

Under inertia free condition, power output is a monotonic decreasing function of frequency as described by Eq.\ \ref{eq:wnores}. The zero frequency asymptotes are $\gamma_m/(\alpha_m^2+\zeta_m)$ for the membrane model and $\gamma_1/(\alpha_1^2+\zeta_1)$ for the 1D model. The ratio of these zero-frequency asymptotes is plotted in Fig.\ \ref{fig:noinertia}A. 

For high frequencies the power output declines proportional to $(1/\omega)^2$. The coefficients are proportional $\gamma^2$. The ratio of the coefficient for the membrane model to that for the 1D model is plotted in Fig.\ \ref{fig:noinertia}B.

These ratios are very close to unity near $K=k$ (blue traces) and deviate significantly for larger elastic load at both ends of the membrane potentials. However, these deviations are not significant near the resting level of the membrane potential (Fig.\ \ref{fig:noinertia}).

The previous analysis based on the 1D model indicates that the optimal elastic load $K$ to for counteracting viscous drag is $K\approx k$ for physiological operating point near $-50$ mV \cite{Iwasa2016}.

The comparison shows that the power output of the membrane model is similar to that of the 1D model in the physiological membrane potential range. Outside of this voltage range, the membrane model predicts smaller power output than the 1D model. This smaller output is also dependent on the elastic load. 

\subsubsection{Near resonance}
\begin{figure}[h!]
\begin{center}
\includegraphics[width=10cm]{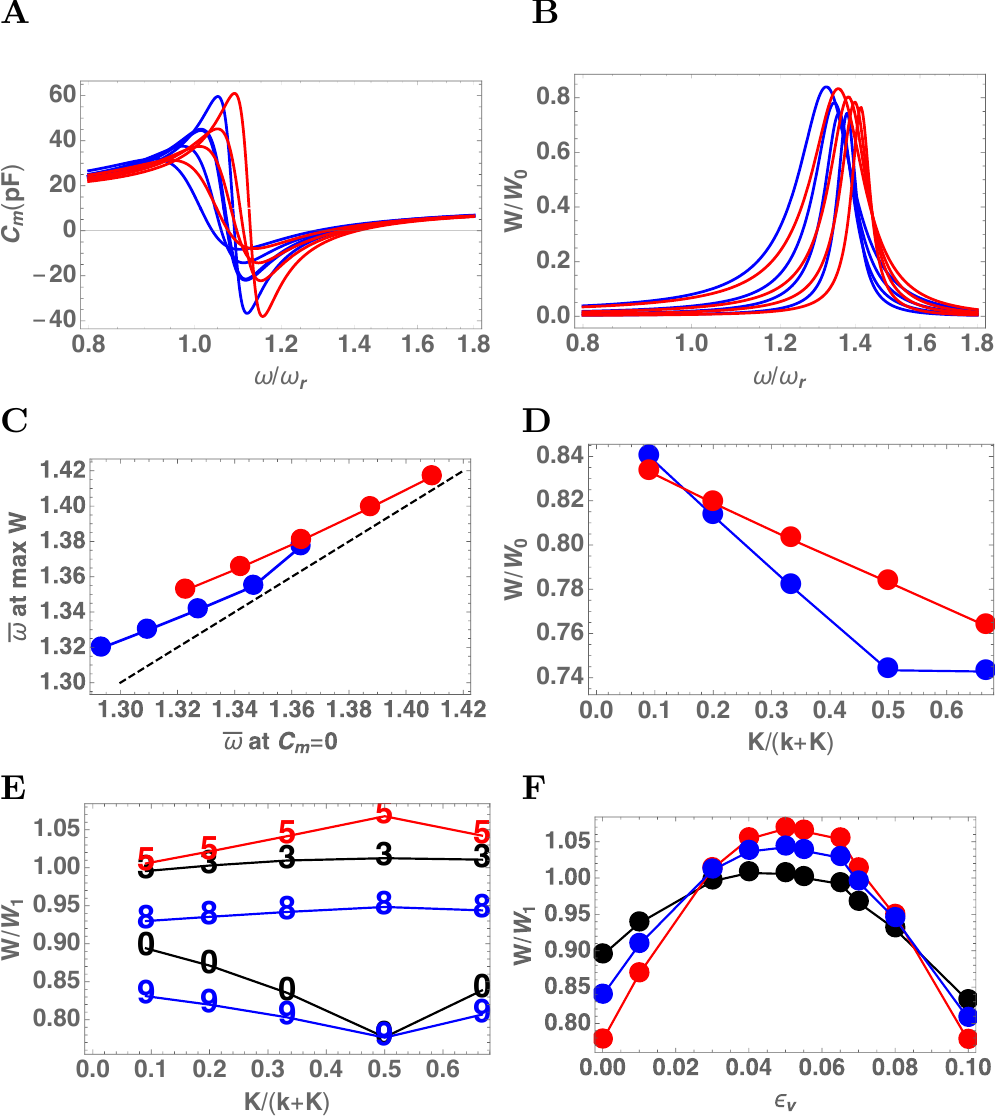} 
\caption{\small{The membrane capacitance and the power output of an OHC. \textbf{A}: The membrane capacitance plotted against the reduced frequency $\omega/\omega_r$. The set of plots are generated by increasing the external elastic load $K$. The values of $K$ correspond to $0.1k, 0.5k, k, \mathrm{and}\; 2k$ (from the left to right). Red:\ membrane model. Blue:\ 1D-model. \textbf{B}: Power output function plotted against frequency. The set of plots are generated by increasing $K$ in a manner similar to A. \textbf{C}: Frequency of maximum power output against the frequency of zero membrane capacitance. The value of $K$ corresponds to from the left to the right, $0.1k, 0.25k, 0.5k, k,$ and $2k$.  The broken line indicates equal frequency. \textbf{D}: Power output based on the membrane model ($W_m$: red) and on the 1D model ($W_1$: blue) plotted against $\hat K(=\!K/(k+K))$.  $W_0=\eta(a_1i_0\hat r)^2/(2\pi C_0)$. \textbf{E, F}: Ratio of power output of the membrane model to that of 1D model.  \textbf{E}: Dependence on the elastic load $\hat K$. The numbers in the ascending order correspond to $\epsilon_v=0$  (labeled 0, black), 0.03 (3, black), 0.05 (5, red), 0.08 (8, blue), and 0.1 (9, blue).  \textbf{F}: Dependence on volume strain $\epsilon_v$. $\hat{K}=0.05$ (red), 0.7(Blue), 0.1(black). In A--D, $\epsilon_v=0.065$. The drag factor $\omega_\eta/\omega_r$ is assumed to be 5 at $\hat K=0.$
}}
\label{fig:wmax}
\end{center}
\end{figure}

It has been shown with the 1D model that nonlinear capacitance is negative near resonance frequency  and can make the total membrane capacitance negative (blue traces in Fig.\ \ref{fig:wmax}A) and that the frequency of maximum power output (blue traces in Fig.\ \ref{fig:wmax}B) is close to the frequency of zero capacitance in such cases (blue traces in Fig.\ \ref{fig:wmax}C).

The membrane model predicts a similar relationship between the membrane capacitance and power output (red traces in Fig.\ref{fig:wmax}A--C  are for $\epsilon_v=0.065$). Power output predicted by the membrane model is slightly smaller than that of the 1D model for small load. However the membrane model predicts larger power output for large load $K$ (Fig.\ \ref{fig:wmax}D).

The difference between the two stems from the definitions of $\Delta G_m$ (Eq.\ \ref{eq:Gm}) and $\Delta G_1$ (Eq.\ref{eq:G1}). For small $K$, the motile response of the membrane model is less sensitive than the 1D model because it has a negative feedback term that does not diminish with the load. However increased load $K$ can reverse the significance of negative feedback because an increase in the load $K$ does not increase negative feedback in the membrane model as much as it does in the 1D model.

The ratio $W_m/W_1$ of power output predicted by the membrane model to that of the 1D model depends on both turgor pressure and the elastic load (Fig.\ \ref{fig:wmax}E, F). The ratio is larger than unity for $\epsilon_v=0.05$ and maximizes at $\hat K=0.5$, where it is about 1.07 (Fig.\ \ref{fig:wmax}E). The ratio is lower at both larger and smaller values of $\epsilon_v$, where ratio decreases between 0.8 and 0.9.

For every fixed value of the elastic load, the power output ratio has a broad maximum at $\epsilon_v=0.05$ (Fig.\ \ref{fig:wmax}F). The dependence on turgor pressure is sharpest for $\hat K=0.5$, i.e. the stiffness of the elastic load is the same as the material stiffness of the OHC (Fig.\ \ref{fig:wmax}F). The ratio is between 1.07 (at $\hat K=0.5$) and about 0.8.

In the previous analysis based on the 1D model,  estimated power output of an OHC was between 0.1 and 10 fW near resonance frequency \cite{Iwasa2017}.   The effectiveness of OHC under \emph{in vivo} condition was estimated by evaluating a limiting frequency at which the power output of the OHC was equal to the viscous loss,  assuming that the major contribution is from the gap between the tectorial membrane and the reticular lamina \cite{allen1980,Wang2016,Hemmert2003}. This frequency is constrained by two factors, resonance frequency and impedance matching: higher resonance frequency requires stiffer elastic load, which leads to poorer impedance matching for power transfer. This evaluation led to the limit of $\sim$10 kHz if OHC is directly associated with the motion of the basilar membrane \cite{Iwasa2017}. To support higher frequencies OHC needs to be associated to smaller elastic load and smaller mass, requiring multiple modes of motion in the organ of Corti \cite{Iwasa2017}. The slightly larger ($\sim$7$\%$) power output predicted by the membrane model leads to a slightly higher limiting frequency. 

\section{Discussion}
It was assumed in the beginning that the cylindrical shape of the cell is maintained while the cell is driven by changes in the membrane potential and undergo deformation. First, the validity of this assumption is examined. A somewhat related issue is the magnitude of the internal drag. That is discussed next. That is followed by possible implications of turgor pressure dependence. In the last part, the comparison with standard piezoelectric resonance is discussed.

\subsection{Limitation of validity}
The conservation of the cylindrical shape of the cell during motion of the OHC requires that the elastic force of the membrane exceeds the inertial force of the internal fluid. Let $x$ the amplitude of the end-to-end displacement of a cylindrical cell of radius $r$ and length $L$. This condition can be expressed by
\begin{align}
2\pi r \kappa\cdot\frac{x}{L}\gg \rho\pi r^2x\cdot\omega^2x,
\end{align}
where $\kappa$ is the elastic modulus of the cell (in the axial direction), $\rho$ the density of the internal fluid, $\omega$ the angular frequency.  The inequality can be expressed by defining a  frequency $\omega_\mathrm{bal}$, at which these two factors are balanced, as
\begin{align}
 \omega\ll \omega_\mathrm{bal} \equiv \sqrt{\frac{2\kappa}{\rho r x L}}.
\end{align}
This expression is intuitive in that a smaller displacement, a decrease in cellular dimension as well as an increase in the elastic modulus favor the elastic force over the inertial force. 

The experimentally obtained value for $2\pi r\kappa$ is 510 nN per unit strain and it is reasonable to use the density of water $10^3$ kg/m$^3$ for the density $\rho$. The radius $r$ is 5 $\mu$m and the length $L$ is 10 $\mu$m for a basal cell. If we assume the amplitude $x$ is 1 nm, an approximate magnitude under \emph{in vivo} condition, the limit can be expressed by the linear frequency
\begin{align}
 f_\mathrm{bal}= 4\times 10^6 \mathrm{Hz},
\end{align}
about 40 times higher than about 100 kHz for high frequency mammals such as bats and dolphins. The condition is even more favorable for more apical cells because $ f_\mathrm{bal}$ decreases with to the square root of $1/L$ whereas the best frequency decreases much steeper.

This means we can reasonably assume that relative motion of the internal fluid against the plasma membrane can be ignored and that the main mode of cell deformation is elongation and contraction while keeping the cylindrical shape.

\subsection{The role of turgor pressure}

The predicted dependence of power output of OHC on turgor pressure (Fig. \ref{fig:wmax}D) raises a number of interesting questions. What is the range of turgor pressure \emph{in vivo}? How much turgor pressure can change? Whether can it function as a control parameter of the cochlear function? 

 Power output has a plateau with respect to $\epsilon_v$ at about $\epsilon_v=0.05$  (Fig. \ref{fig:wmax}D), which corresponds to a static axial strain of $\epsilon_z=-0.018$ under load-free condition. This strain is  about 36\% of the maximum amplitude of electromotility (5\% of the cell length). The corresponding turgor pressure is 0.14 kPa.  It is probable that the physiological turgor pressure could be lower than this value as often the case for biological functions. Such an operating condition allows gain control by the parameters. 

 However, turgor pressure is not a simple control parameter of power output because changes in turgor pressure accompanies shifts of the operating voltage (see e.g. Fig.\ \ref{fig:cap}) even though its effect on the maximal power output is to up 20\% (Fig.\ \ref{fig:wmax}F). 

\subsection{Piezoelectric resonance}\label{subsec:piezo}
The derivation of the equation of motion (Eq.\ \ref{eq:eom}) may not appear legitimate in that it introduces the inertia term to a stochastic equation. However, it turns out to be consistent with a standard expression for the admittance of a piezoelectric system.

The standard expression for the admittance $Y_\mathrm{pe}$ of a piezoelectric resonator can be \cite{i1990}
\begin{align}
Y_\mathrm{pe}(\omega)=i\omega C_0+\frac{1}{R+i[\omega L_p -1/(\omega C_p)]},
\end{align}
using an equivalent electric circuit  with inductance $L_p$, and resistance $R$.  That implies the correspondence to mechanical resonance system: $\omega_r^2=1/(C_pL_p)$ and $\omega_\eta=1/(R C_p)$, leading to
\begin{align}
 Y_\mathrm{pe}(\omega)=i \omega C_0+\frac{i C_p}{i \omega/\omega_\eta+1-(\omega/\omega_r)^2}.
 \label{eq:Ype}
\end{align}
Eq.\ \ref{eq:Ype} is equivalent with Eq.\ \ref{eq:cmemb_eq} because $Y_{nl}=i\omega Nqc$ and the zero-frequency limit indicates $C_p=\gamma nq^2$ and $\alpha=1$, which corresponds to $K=0$ since the external spring does not exist for the piezoelectric element.

This comparison also illustrates a limit of validity for the equation of motion (Eq.\ \ref{eq:eom}). While Eq.\ \ref{eq:Ype} for standard piezoelectricity does not depend on the operating point, Eq.\ \ref{eq:cmemb_eq} for OHC does through the linearization near the equilibrium condition though the factor $\gamma$ (=$\beta \langle C\rangle(1-\langle C\rangle))$. The equation of motion Eq.\ \ref{eq:eom} is valid only within a small range of the membrane potential, in which linearization can be justified. 

\section{Conclusions}

The membrane model predicts nonlinear capacitance, cell displacement, and power output of OHCs relevant to \emph{in vivo} conditions. In addition, these predictions are testable by \emph{in vitro} experiments.

Nonlinear capacitance is sensitive to both turgor pressure and external elastic load. An increased elastic load reduces the peak hight and broadens the voltage dependence of nonlinear capacitance. 
The peak voltage shifts in the positive direction with increasing turgor pressure and elastic load. That is intuitive because increasing internal pressure positively shifts the capacitance peak.

Power output depends on turgor pressure. The optimal power output is expected at $\epsilon_v=0.05$ and $\hat K=0.5$. Under this condition, maximal power output is about 7 \% higher than the previous estimate based on the 1D model. However, the dependence of power output on turgor pressure is not large. The deviations from the predictions of 1D model do not exceed 20 \%.

The membrane model confirms the main predictions of the 1D model: A single mode of vibration of organ of Corti can be supported up to about 10 kHz but to cover the entire auditory range cannot be supported without multiple modes of motion in the cochlear partition \cite{Iwasa2017}. This prediction appears consistent with recent observations with optical techniques in that the organ of Corti shows multiple modes of motion \cite{Gao2014,He2018,Cooper2018}. 

\section*{Acknowledgments}
The author thanks Dr.\ Richard Chadwick for discussion. This research was supported in part by the Intramural Research Program of the NIH, NIDCD.


\pagebreak

\numberwithin{equation}{section} 
\numberwithin{figure}{section}

\appendices 
\section{Derivations of $\epsilon_ z$ and $\Delta G_m$}\label{apx:eqs}
 The constitutive equations are given by Eqs.\ \ref{eq:total_strains}:
\begin{align*}
d_1\epsilon_z+g\epsilon_c-(a_zd_1+a_cg)nC&=f_z+\frac 1 2 rP,\\
g\epsilon_z+d_2\epsilon_c-(a_zg+a_cd_2)nC&= rP.
\end{align*}
For a cylindrical cell of length $L$ and radius $r$, the cell volume is given by $V=\pi r^2L$. For small length strains, $\epsilon_z$ in the axial direction and $\epsilon_c$ in the circumferential direction, the volume strain $\epsilon_v$ can be expressed by 
\begin{align}\label{eq:volume}
 \epsilon_v=\epsilon_z+2\epsilon_c. 
\end{align}
By eliminating the circumferential strain $\epsilon_c$ from these equations, we obtain
\begin{subequations}\label{eq:rewrite}
\begin{align} \label{eq:2d1} 
(2d_1-g)\epsilon_z+g\epsilon_v-2(a_zd_1+a_cg)nC& = 2f_z+rP,\\ \label{eq:gd2} 
(g-\frac {d_2} {2} )\epsilon_z+\frac {d_2} {2} \epsilon_v-(a_zg+a_cd_2)nC& = rP.
\end{align}
\end{subequations}
 By eliminating $rP$ from Eqs.\ \ref{eq:rewrite}, an expression for $\epsilon_z$ can be obtained. Then by replacing $\epsilon_z$ in Eq.\ \ref{eq:gd2} with this expression, we obtain an expression for $rP$. They are 
\begin{subequations}
\label{eq:with_fz}
\begin{align}\label{eq:zfz}
2\kappa\epsilon_z&=-AnC+\mu \epsilon_v+2f_z,\\ \label{eq:2rpfz}
2\kappa rP&=-\varphi a nC+\varphi \epsilon_v-2\mu f_z,
\end{align}
\end{subequations}
with short-hand notations
\begin{subequations}
\begin{align*}
A&=-\nu a_z+2\mu a_c, &\mu=d_2/2-g,\\
a&=-(a_z+2a_c), &\nu=2d_1-g,\\
\varphi&=d_1d_2-g^2, &\kappa=d_1+d_2/4-g.
\end{align*}
\end{subequations}
Notice here that the parameters $A$ and $a$ are defined such that they are positive.
In the absence of the motile elements in the membrane Eq.\ \ref{eq:zfz} is reduced to $f_z=-\kappa\epsilon_z$. This implies that $\kappa$ is the axial elastic modulus. See Eqs.\ \ref{eq:k_kappa} and \ref{eq:kappa}.

In the presence of external elastic load, $f_z=-K_e\epsilon_z$. Then Eq.\ \ref{eq:zfz} turns into
\begin{align}\label{eq:app-a2kap}
2(\kappa+K_e)\epsilon_z&=-AnC+\mu \epsilon_v,
\end{align}
where $\hat{K}=K_e/(\kappa + K_e)$. By substituting $f_z$ in Eq.\ \ref{eq:2rpfz} with $-K_e\epsilon_z$ using Eq.\ \ref{eq:app-a2kap}, we obtain
\begin{align}\label{eq:app-a2kap}
2\kappa rP&=-(\mu A\hat{K}+\varphi a)nC+(\mu^2\hat{K}+\varphi)\epsilon_v.
\end{align}
Notice that Eq.\ \ref{eq:app-a2kap} is the same as Eq.\ \ref{eq:eps_z1} in the main text.

With the aid of Eq.\ \ref{eq:app-a2kap}, the axial stress $f_z$ can be expressed as
\begin{align}\label{eq:fz_final}
 f_z=\frac1 2 \hat{K}(AnC-\mu \epsilon_v).
\end{align}
The free energy $\Delta G_m$ of state C referenced from state E is given by 
\begin{align}\label{eq:DeltaG}
 \Delta G_m=-q(V-V_0)-a_zf_z-(a_c+a_z/2)rP,
\end{align}
where $qV_0$ is a constant term. Since $q<0$ and $a_c+a_z/2<0$ \footnote{Unlike previous treatments, the state E is chosen here as the reference for the convenience of extension of this treatment to multi-state models.}, both depolarization and increased turgor pressure lead to a decrease of state C.
By substituting $rP$ and $f_z$ in Eq.\ \ref{eq:DeltaG} respectively with Eqs.\ \ref{eq:app-a2kap} and \ref{eq:fz_final}, we obtain
\begin{align}
 \Delta G_m=-q(V-V_0)+\frac{1}{4\kappa}[(A^2\hat{K}+\varphi a^2)nC-(\mu A\hat{K}+\varphi a)\epsilon_v],
\end{align}
which is Eq.\ \ref{eq:DGm} in the main text.

\section{Internal drag}\label{apx:i-drag}

For frequencies that satisfy $\omega\ll\omega_\mathrm{char}$, the shape of an OHC can be approximated by a cylinder and the displacement of the cell could be virtually determined by the membrane elasticity alone as under static condition. 

\begin{figure}[h!]
\begin{center}
\includegraphics[width=5cm]{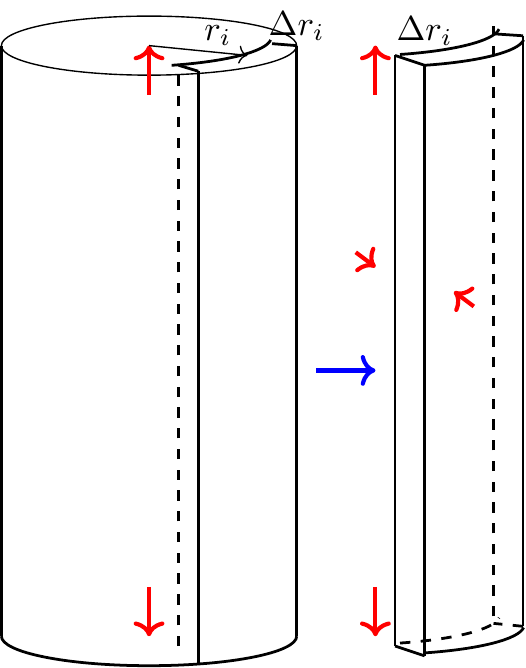} 
\caption{\small{Displacement of intracellular fluid. }}
\label{fig:int_drag}
\end{center}
\end{figure}

Consider a part of a cylindrical layer of thickness $\Delta r_i$, the inner surface of which is located at distance $r_i$ from the center. While the cell is elongating, the center of gravity moves toward inside due to constant volume condition, the internal fluid being incompressible. This layer elongates uniformly in the axial direction without slippage. The outer border moves more than the inner border does in the radial direction, but the direction of the movement is perpendicular to the surface and does not contribute to viscous drag.  This description applies throughout the cell interior. During the shortening of the cell, the reverse movement likewise does not involves slippage. For this reason, viscous drag must be virtually absent inside of an OHC,  except for the lateral wall, the vicinity of the nucleus and the apical plate. 

\section{Drag at the external surface}\label{apx:e-drag}
\subsection{in vitro condition}\label{app:in_vitro}
Here we assume the roll-off frequency $\omega_{f}$ observed under load-free condition is determined by the drag at the external surface of OHCs and the cell's stiffness. Since it is expected that the drag coefficient $\eta_{e}$ at the external surface increases with the length $L$ of the cell exposed to the external fluid, it should be an increasing function of $L$.

The roll-off frequency $\omega_{f}$ under load-free condition is expressed by $\omega_{f}=k/\eta_{e}$. The axial stiffness of the cell $k$ is related to the axial elastic modulus $\kappa$ by $k=\kappa/L$. Thus we obtain an expression for drag coefficient:
\begin{align}\label{eq:eta_e}
\frac{\kappa}{\omega_{f} L} =\eta_e.
\end{align}

\begin{figure}[h!]
\begin{center}
\includegraphics[width=7cm]{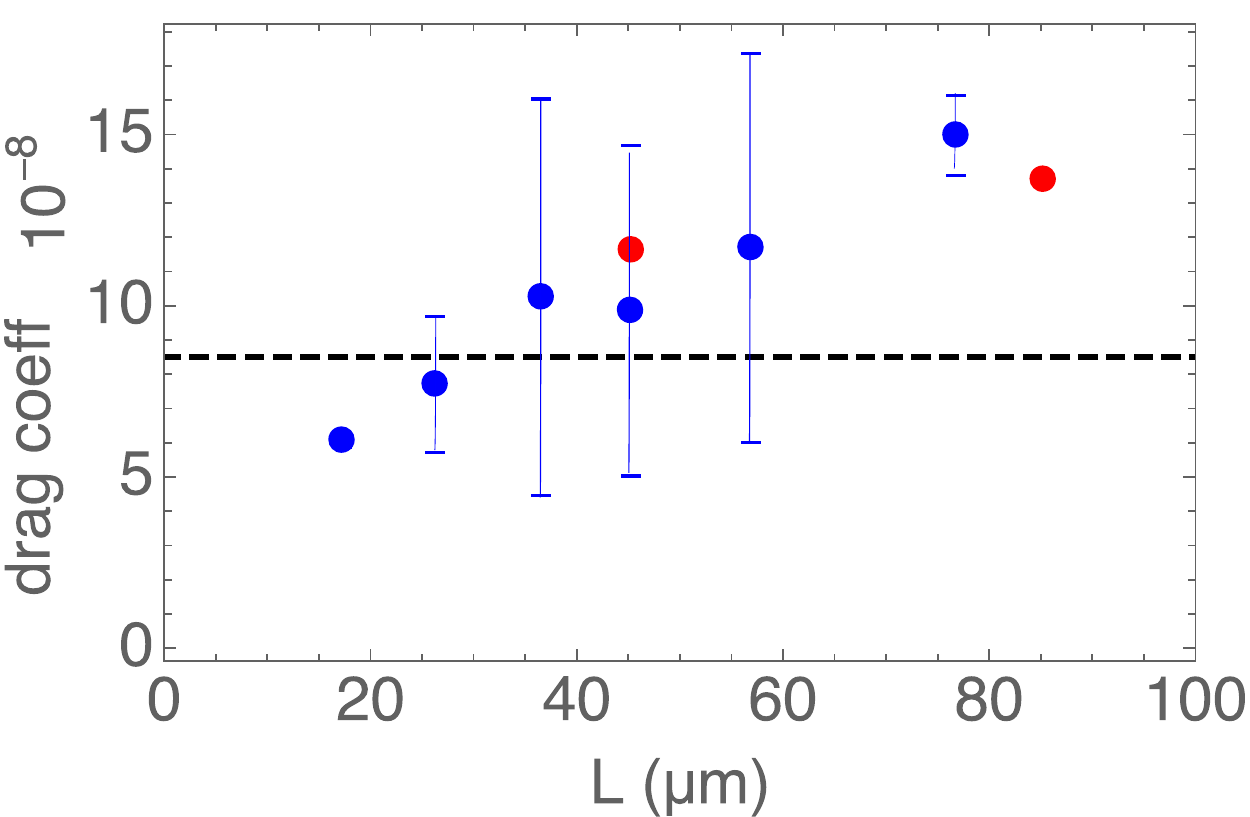} 
\caption{\small{Drag coefficient of the external surface of OHC. The drag coefficient $\eta_e$ (in N$\cdot$s/m) is plotted against cell length L. The value of $\kappa$ is 510 nN per unit strain \cite{ia1997}. The data points are taken from Fig.\ 3A in ref.\ \cite{fhg1999}. The two date points in red are taken from the same cell by changing the ratio of the cell body in the pipette. Error bars show standard deviations. Dashed line indicates the level of the Stokes drag of a sphere of 5$\mu$m radius.}}
\label{fig:ext_dag}
\end{center}
\end{figure}

Eq.\ \ref{eq:eta_e} can be examined with experimental roll-off frequency obtained in the microchamber configuration \cite{fhg1999}. The order of magnitude of the drag coefficient is the Stokes drag of a 5$\mu$m sphere. In addition, the drag coefficient $\eta_e$ appears to be an increasing function of the length $L$ of the cell outside of the pipette, as expected.

This result is consistent with the hypothesis that conformational transitions of prestin is determined by mechanical constraints. It is also consistent with the analysis that internal drag is not significant.

\subsection{\emph{In vivo} condition}\label{apx:in_vivo}
The space of Nuel, which surrounds the basolateral membrane of OHCs, undergoes displacement while OHCs move. If the displacement of this space parallels that of OHCs, an argument similar to the internal friction of OHCs could be made for the external friction. However, that is not the case for a number of reasons. 

This space has structural elements other than OHCs. Deiters' cells, which connects OHCs to the basilar membrane, not only do not undergo active displacements the same as OHCs, their stiffness may not be the same. In addition, the phalangeal processes, which have stiff microtubule backbones, runs at an angle to OHCs, likely twisting the space when OHCs undergo displacement. Moreover, this space is continuous from the base to the apex, allowing fluid flow. Since the synchrony of OHC movement  along the lateral axis should have a limited range, the volume of the extracelluar space in a given segment may not be conserved. 

For these reasons, an analogy to the internal surface does not apply to the external surface for the fluid motion. However, this drag may not be as large as under \emph{in vitro} conditions.

\end{document}